\documentclass[showpacs,preprintnumbers,amsmath,amssymb]{revtex4}\usepackage{graphicx}
\usepackage{dcolumn}
\usepackage{bm}
\usepackage{hyperref}

\begin{document}
\preprint{}
\title{The entropy of Taub-Bolt-AdS from an improved action principle}
\author{Rodrigo Aros}
\affiliation{Departmento de Ciencias Fisicas\\ Universidad Andres Bello, Republica 252,
Santiago,Chile}
\date{\today}
\pacs{04.70.Dy, 04.70.Bw}
\begin{abstract}
In this article the study of four dimensional spaces with NUT charges and their ground states is continued. In particular, through the introduction of an improved action principle containing topological densities, it is shown that these spaces have simple and sound thermodynamical relations. As an example the entropy of Taub-Bolt-AdS solution is computed in terms of a Noether charge.
\end{abstract}

\maketitle

\section{Introduction}

In general terms, within a family of solutions characterized by a set of parameters, a ground state can be identified with a solution in the family whose number of symmetries is larger due to a subset of those parameters vanish, or satisfy a particular condition that reduces the dimension of the space of parameters. For gravity this usually leads to constant curvature spaces or identified constant curvature spaces without fixed points \footnote{See \cite{Wolf1984} for definition of these spaces}. Now, it is tempting to try to identify those parameters with conserved charges, however, this is not general as that requires to be able to define conservation in the space. For that the space must have, at least asymptotically, a time like symmetry.

Once conserved charges can be defined it becomes necessary to identify them. Among the different approaches to identify the conserved charges the Noether method definitively has a prominent role. It is worth to recall that the Noether charges are constructed out of symmetries of the solutions or equivalently from the subset of the symmetries of the theory with parameters satisfying some \emph{Killing conditions}. In the case of gravity the symmetry of interest is the invariance under diffeomorphisms and the \emph{Killing} conditions define the \emph{Killing} vectors.

It is usually overlooked that the soundness of the Noether charges relay on the action principle considered, and thus some mismatches might arise. For instance, the Noether charges computed out of the plain Einstein-Hilbert action, sometimes called Komar charges, have two major drawbacks. Different normalizations are required to compute the mass and the angular momenta of the solutions. They also diverge for asymptotically AdS spaces. In the context of AdS/CFT conjecture this is certainly something that needs to be addressed. For instance, the Holographic Renormalization Method \cite{Skenderis:2002wp}  handles those divergences by the addition of boundary terms that preserve Dirichlet boundary conditions for the metric.

Finally, in the context of this work, it is worth to mention that the first law of thermodynamics for stationary black holes can be also obtained in terms of Noether charges as well as an expression for their entropy \cite{Wald:1993nt}.

The other version of the problem of charges corresponds to the Hamiltonian approach, where the on-shell value of the generators give rise to the value of conserved charges. In \cite{Regge:1974zd} was shown that Hamiltonian generators of diffeomorphisms have two parts, a bulk part and boundary part. The bulk part always vanishes on shell, and actually is only the value (on shell) of the boundary part which is responsible for the value of the charges (such as the mass or angular momenta of the solution). The problem of different normalizations for mass and angular momenta is absent in the Hamiltonian approach. This discrepancy, between the Komar and its Hamiltonian counterpart, was clarified in \cite{Lee:1990nz} using phase space technics. The divergences, mentioned above, for asymptotically AdS spaces can be tamed at the expense of the introduction of an ah-hoc background. In \cite{Hawking:1983dh} a different solution, based on the exterior curvature of the \emph{infinity}, was proposed.

In \cite{Aros:1999id} was proposed a different approach to the problem of charges and regularization of the action principle. This is based on the addition of topological densities and provides a regularized and well defined action principle for asymptotically locally AdS spaces and simultaneously sound Noether charges. Moreover, these Noether charges are in one on one correspondence with their Hamiltonian (charges) counterpart \cite{Aros:2005by}, unlike those constructed out of the Komar potentials. Finally, the Noether charges (constructed out of this new action principle) coincide \cite{Aros:2002ub} with the (generalization of) Brown-York boundary charges defined in \cite{Brown:1992br}. Remarkably there is natural way to extend this approach to any higher even dimension \cite{Aros:1999kt}. This was extended even further in \cite{Olea:2005gb,Miskovic:2009bm}, where it is also shown its equivalence with the holographic renormalization method for all the known cases.

\subsection*{Topological charges as Noether charges}

To fix some ideas one can recall the case of electromagnetism. The action principle $\int F\wedge ^* F$, through the gauge symmetry $A \rightarrow A + d\lambda$, gives rise
\[
  Q(\lambda) = \int_{\partial \Sigma_{\infty}} \lambda ^{*}F
\]
where $\partial\Sigma_{\infty}$ is the radial infinity of a constant time slide of the space. To identify the electric charge as a Noether charge the parameter of the gauge transformation $\lambda$ must satisfies $d\hat{\lambda}=0 \leftrightarrow \hat{\lambda} = const.$, which corresponds to the \emph{Killing} condition for $U(1)$. The electric charge is given by $q_e = Q(\hat{\lambda})/\hat{\lambda}$.

In four dimensions the presence of a non-trivial $U(1)$ fiber bundle determines the presence of magnetic charges whose value can be obtained from
\[
q_m = \int_{\Sigma_{\infty}}  F.
\]
Remarkable, this charge can be incorporated into a Noether charge provided the Pontryagin density, $\int F\wedge F$, is added to the action principle. Notice that this does not change the equations of motion. By considering
\[
I = \int F \wedge ^* F + \beta \int F \wedge F,
\]
the integral of Noether current is given by
\[
Q(\lambda) =   \int_{\Sigma_{\infty}}  \lambda ( ^* F + \beta F),
\]
and therefore $Q(\hat{\lambda})/\hat{\lambda} = q_e + \beta q_m$.

A ground state can be defined by vanishing of the both electric and magnetic charges, \emph{i.e.}, $q_e=0$ and $q_m=0$. However, this is not the only possibility. It can be noticed that by fixing $\beta=\pm 1$ the action can be rewritten as
\[
I =  \pm \frac{1}{2}\int (^*F  \pm F)\wedge (^*F \pm F)
\]
and therefore for the (anti) self dual case, $^*F = \pm F$ the action vanishes. Furthermore, considering Euclidean version of the action principle, this has its minimum value in this case. This also defines $Q(\hat{\lambda})=0$. With this in mind, any (anti-)self dual solution can be casted a \emph{ground state} as well.

In \cite{Araneda:2016iiy} the ideas above were extended to gravity in terms of the separation of the Weyl tensor into its electric and magnetic parts. There it was proposed and justified the idea that spaces whose associate Weyl tensors are non trivially (anti) self dual can be casted as proper ground states. For this an improved action principle is introduced. This action is regularized for asymptotically locally AdS spaces with topological defects and the associated Noether charges vanish for spaces whose Weyl tensor is (anti) self dual.

In four dimensions the simplest non-trivial examples whose Weyl tensor is (anti) self dual are the Taub-NUT and Taub-NUT-AdS solutions. These solutions contain what is usually called a NUT charge, see for instance \cite{Hunter:1998qe}, but still are asymptotically locally flat and asymptotically locally AdS respectively. For the AdS/CFT conjecture this opened the possibility to address CFT's defined on conformal manifolds with topological non-trivial defects. For a discussion about this see \cite{Chamblin:1998pz}.

This article aims to proceed with the analysis of the action principle presented in \cite{Araneda:2016iiy}. Here it is shown that its Noether charges are in one-on-one correspondence with their Hamiltonian counterpart. It is also computed, as an example, the entropy of the Taub-Bolt-AdS solution in terms of a Noether charge.

Before to proceed it is worth to stress that the spaces to be considered in this work are Euclidean and have a well defined asymptotically locally AdS region. In general these spaces it will considered to have a well defined split as $\mathcal{M}=S^1\times \Sigma$ where $\Sigma$ corresponds to a 3-dimensional spacelike hypersurface and $\mathbb{R}$ stands for the time direction. In addition, $\partial\Sigma$ will be considered the union of an exterior and an interior surface, thus $\partial \Sigma=\partial\Sigma_{\infty} \oplus \partial \Sigma_{H}$.

To keep the notation as close as possible to a gauge theory the computation will be expressed in first order formalism. However, the pass to second order formalism is direct.

\section{Addition of topological invariants densities and charges}

Let's start by reviewing the original proposal in four dimensions \cite{Aros:1999id}. This corresponds to the addition to the Einstein Hilbert action (plus the negative cosmological constant) of the Euler density. This allows to express the action principle, either in first or second order, as
\begin{eqnarray}
  I_{reg} &=& \frac{l^2}{64\pi G} \int \bar{R}^{ab}\bar{R}^{cd}\varepsilon_{abcd}\label{RRaction}  \\
   &=& \frac{l^2}{64\pi G} \int \delta^{\mu_1\mu_2\mu_3\mu_4}_{\nu_1\nu_2\nu_3\nu_4}\bar{R}^{\nu_1\nu_2}_{\hspace{2ex}\mu_1\mu_2}\
  \bar{R}^{\nu_3\nu_4}_{\hspace{2ex}\mu_3\mu_4} \sqrt{g} d^4 x \nonumber
\end{eqnarray}
where
\begin{equation}\label{Rbar}
  \bar{R}^{\nu_1\nu_2}_{\hspace{2ex}\mu_1\mu_2} = R^{\nu_1\nu_2}_{\hspace{2ex}\mu_1\mu_2} + \frac{1}{l^2} \delta^{\mu_1\mu_2}_{\nu_1\nu_2}
\end{equation}
with $R^{\nu_1\nu_2}_{\hspace{2ex}\mu_1\mu_2}$ is the Riemann tensor. The cosmological constant is given by $\Lambda = -3 l^{-2}$. In Eq.(\ref{RRaction})
\[
R^{ab} = \frac{1}{2} e^a_{\nu_1}e^b_{\nu_2}  R^{\nu_1\nu_2}_{\hspace{2ex}\mu_1\mu_2} dx^{\mu_1} \wedge dx^{\mu_2}
\]
is called the curvature two-form. Here $\{e^{a}_{\hspace{1ex}\nu}\}$ is a orthonormal basis of four dimensional (co-)vectors of the co tangent space of the manifold $\mathcal{M}$. This defines a vielbein  $e^a = e^{a}_{\hspace{1ex}\nu} dx^{\nu}$.

To continue with the discussion one can observe that the action principle above (\ref{RRaction}) vanishes for any locally AdS space. In this way, this action principle is tailored such that any locally AdS spaces can be casted as a ground state. Moreover, this is finite for any asymptotically locally AdS solution.

On remarkable fact about the action principle (\ref{RRaction}) is that on shell it can be casted as
\begin{equation}\label{WeylGravity}
\left. I_{reg} \right|_{\textrm{on shell}} = \frac{l^2}{64\pi G} \int C^{\nu_1\nu_2}_{\hspace{2ex}\mu_1\mu_2} C_{\hspace{2ex}\nu_1\nu_2}^{\mu_1\mu_2} \sqrt{g} d^4x,
\end{equation}
where $C^{\nu_1\nu_2}_{\hspace{2ex}\mu_1\mu_2}$ is the Weyl tensor. This connects this action principle with Weyl Gravity in the same fashion as was explored in \cite{Maldacena:2011mk}. In this context it is worth to mention \cite{Miskovic:2009bm} where the connection between Eq.(\ref{WeylGravity}) and the standard electromagnetic action, $\int F^{*}F$, was original discussed in this terms.

In this moment is worthwhile to mention that in \cite{Aros:1999id} was discussed the value of the mass of Taub-Nut-AdS and Taub-Bolt-AdS solutions. This was done in the context of the previous results by Hawking \cite{Hawking:1998ct} where the mass of Taub-Bolt-AdS was computed with respect to the Taub-Nut-AdS solution. The result in \cite{Aros:1999id} showed that both masses can be computed separately and that actually their difference reproduces the result in \cite{Hawking:1998ct}. Because of that, the mass computed out of the action principle Eq.(\ref{RRaction}) can be casted as the electric $mass$ of the solution. See \cite{Araneda:2016iiy}.

\subsection*{On gravitational solitons}

In \cite{Araneda:2016iiy} the previous analysis was continued for spaces whose Weyl tensor is (anti-)self dual. In this section these ideas are re-expressed in first order formalism. The Einstein Hilbert action is not only regularized for asymptotically locally AdS spaces, by the addition of Euler density, but also complemented by the addition of the Pontryagin density such that
\begin{equation}\label{RPaction}
  I_{GP} = I_{reg} +  \alpha \int R^{ab} R_{ab}
\end{equation}
where $\alpha$ is constant. This term can be identified with $\int F \wedge F$ for electromagnetism.

As discussed in details in \cite{Araneda:2016iiy} the addition of Pontryagin density, though in principle alters the value of the action principle, does not modify the finiteness of the action principle. Moreover, it must be stressed that most of the renown analytic solutions, such as Schwarzschild or Kerr AdS, have vanishing Pontryagin density. Therefore, any ground state of the those solutions can be casted as ground state of the action principle Eq.(\ref{RPaction}). This is in complete analogy to electromagnetism where, in absent of magnetic charges, the ground state is defined by the vanishing of the electric charge.

It must be noticed that, due to the solutions are Einstein manifolds, that
\[
 R^{ab} R_{ab} = \bar{R}^{ab} \bar{R}_{ab}.
\]

It is worthwhile to notice the presence of a AdS pedigree in this case. Provided one defines a AdS$_4$ connection in terms of the vielbein and the spin connection as $W^{AB}=(\omega^{ab},e^a/l)$, see for instance \cite{Zanelli:2002qm}, therefore AdS$_{4}$ Pontryagin density $\mathcal{F}^{AB}\mathcal{F}_{AB}$ can be expressed as
\[
\mathcal{F}^{AB}\mathcal{F}_{AB}=\bar{R}^{ab} \bar{R}_{ab} - 2T^a
\]
where $T^a$  is the torsion two form. In the case at hand, $T^a=0$ and therefore one has the identity
\[
\mathcal{F}^{AB}\mathcal{F}_{AB} = \bar{R}^{ab} \bar{R}_{ab}.
\]

Now one can return to address how to define a ground state for solutions with non vanishing Pontryagin $\bar{R}^{ab} \bar{R}_{ab}$ density. This arises by observing that
\[
\left.\int R^{ab} R_{ab}\right|_{\textrm{on shell}} = \left.\int \bar{R}^{ab} \bar{R}_{ab}\right|_{\textrm{on shell}} = 4 \int (C)^{\nu_1\nu_2}_{\hspace{2ex}\mu_1\mu_2} (C^*)_{\hspace{2ex}\nu_1\nu_2}^{\mu_1\mu_2}\sqrt{g} d^4 x
\]
where $(C^*)_{\hspace{2ex}\nu_1\nu_2}^{\mu_1\mu_2} = \frac{1}{2} C_{\nu_1\nu_2\alpha\beta}\varepsilon^{\alpha\beta\mu_1\mu_2}
$ can be identified with the dual of the Weyl tensor.  The dual of the $\bar{R}^{a}b$ is given by $(\bar{R}_{ab})^* = \frac{1}{2}\varepsilon_{abcd}\bar{R}_{cd}$.

With this in mind the $\alpha$ constant in Eq.(\ref{RPaction}) can be fixed by requiring that the action principle satisfies (on shell)
\[
\left.I_{GP}\right|_{\textrm{on shell}} \backsim \int (C \pm C^*)^2.
\]
This is made in order to identify as the ground state any solution with (anti) self dual Weyl tensor.

With $\alpha=\pm 1$ the solutions with (anti) self dual Weyl tensor can be considered, in a broad sense, as \emph{instantons} of a conformal theory of gravity. On the other hand, this choice of $\alpha$ give rise to a natural extension of the previous action principle Eq.(\ref{RRaction}) whose ground states are locally AdS manifolds, and thus trivially have self (anti) dual Weyl tensors. Finally, in first order formalism, the action principle is given by
\begin{equation}\label{RPactionFinal}
  I_{GP} = I_{reg} \pm  \frac{l^2}{32\pi G} \int \bar{R}^{ab} \bar{R}_{ab}.
\end{equation}

To finish the section by noticing that one can define $F^{ab} = \bar{R}^{ab} \pm \frac{1}{2}\eta^{aa'}\eta^{bb'}\varepsilon_{a'b'cd}\bar{R}^{cd}$ such that
\[
F^{ab}F_{ab}=(\bar{R}^{ab} \pm (\bar{R}^{ab})^* )(\bar{R}_{ab} \pm (\bar{R}_{ab})^* )= (\bar{R}^{ab}\bar{R}^{cd}\varepsilon_{abcd} \pm 2 \bar{R}^{ab}\bar{R}_{ab}),
\]
which allows to rewrite action principle as
\[
I_{GP} = \frac{l^2}{64\pi G} \int F^{ab}F_{ab}
\]

\section{Hamiltonian Charges versus Noether Charges}

In \cite{Araneda:2016iiy} was shown that the Noether charges associated with the Killing vectors are given by
\begin{equation}\label{NotherCharges}
  Q(\xi) = \frac{l^2}{32\pi G} \int_{\partial \Sigma} I_{\xi} \omega^{ab}(\bar{R}^{cd} \varepsilon_{abcd} \pm 2 \bar{R}_{ab}).
\end{equation}
As mentioned above the Noether charges of the action principle Eq.(\ref{RRaction}) are exactly the Hamiltonian charges \cite{Aros:2005by}. In order to analyze this for the Noether charge Eq.(\ref{NotherCharges}), defined from the action principle Eq.(\ref{RPactionFinal}), is enough to follow \cite{Lee:1990nz}. It is direct to demonstrate that the variation along the parameters, say $\hat{\delta}$ \cite{Lee:1990nz}, of the Hamiltonian generator associated a diffeomorphisms defined by $x \rightarrow x + \xi$ is given on shell by
\begin{widetext}
\begin{equation}\label{VariationOfHamiltian}
  \left.\hat{\delta} H(\xi) \right|_{\textrm{ on shell}} = \hat{\delta} G(\xi) =  \hat{\delta} ( Q(\xi)) + \int_{\partial \Sigma_{\infty} \oplus \partial \Sigma_{\mathcal{H}}} I_{\xi} \left(\hat{\delta}\omega^{ab}(\bar{R}^{cd} \varepsilon_{abcd} \pm \bar{R}_{ab})\right)
\end{equation}
\end{widetext}
One can recognize that the contribution from $\partial \Sigma_{\infty}$ of the second term vanish for any asymptotically (anti) self dual Weyl space. To address the internal boundary, $\partial \Sigma_{\mathcal{H}}$,  is necessary to impose boundary conditions. As discussed in \cite{Aros:2005by} $\left.\delta\omega^{ab}\right|_{\partial \Sigma_{\mathcal{H}}} = 0$ is indeed a proper boundary condition and therefore the Noether charges of the action Eq.(\ref{RPactionFinal}) can be fixed such as
\[
G(\xi) =   Q(\xi),
\]
proving that the Noether charges Eq.(\ref{NotherCharges}) are indeed equivalent to the Hamiltonian charges.

It must be noticed that the boundary condition $\left.\delta \omega^{ab}\right|_{\partial \Sigma_{\mathcal{H}}} = 0$ fixes the temperature in the case that $\partial \Sigma_{\mathcal{H}}$ is connected with presence of a Killing horizon defined by $\xi$. To observe that one can notice that the temperature of the Killing horizon can be read from the relation \footnote{ It is direct to demonstrate that  Eq.(\ref{Temperature}) is equivalent to the relation $\xi^{\mu}\nabla_{\mu}(\xi^{\nu})|_{\mathbb{R}\times
\partial\Sigma_{H}} = \kappa \xi^{\nu}$ discussed in \cite{Wald:1993nt}.}
\begin{equation}\label{Temperature}
I_{\xi} \omega^{a}_{\hspace{1ex} b} \xi^{b}|_{\mathbb{R}\times
\partial\Sigma_{H}} = \kappa \xi^{b},
\end{equation}
where $\kappa$ is the surface gravity. The temperature is given by $T=\kappa/4\pi$.

\section{Entropy in a Noether charge}

To obtain the thermodynamics defined by the action principle in Eq.(\ref{RPactionFinal}) one can follow the ideas in \cite{Wald:1993nt}. As a detailed discussion about this in first order formalism can be found in \cite{Aros:2002ub}, only the highlights will be discussed here. To obtain the first law is enough to notice, see \cite{Lee:1990nz} for the general expression, that the conservation of the flux along trajectories defined by the variations of parameters of the solutions, within the space configurations, implies that
\begin{widetext}
\[
   \hat{\delta} \frac{l^2}{32\pi G} \int_{\partial \Sigma_{\mathcal{H}}} I_{\xi} \omega^{ab}(\bar{R}^{cd} \varepsilon_{abcd} \pm 2 \bar{R}_{ab}) = \hat{\delta} \frac{\beta l^2}{32\pi G} \int_{\partial \Sigma_{\infty}} I_{\xi} \omega^{ab}(\bar{R}^{cd} \varepsilon_{abcd} \pm 2 \bar{R}_{ab}).
\]
\end{widetext}
Next, one can observe that RHS corresponds to variation of the Noether charges evaluated in $\partial \Sigma_{\infty}$, such as the mass and the rest of the conserved charges,\emph{ i.e.},
\begin{equation}\label{FirstLaw}
  \hat{\delta} \left(\frac{l^2}{32\pi G} \int_{\partial \Sigma_{\mathcal{H}}} I_{\xi} \omega^{ab}(\bar{R}^{cd} \varepsilon_{abcd} \pm 2 \bar{R}_{ab})\right) = \hat{\delta} M + \ldots
\end{equation}
which allows to identify, due to the $T\hat{\delta} S = \hat{\delta} M + \ldots$,
\[
T \hat{\delta} S = \hat{\delta} \left(\frac{l^2}{32\pi G} \int_{\partial \Sigma_{\mathcal{H}}} I_{\xi} \omega^{ab}(\bar{R}^{cd} \varepsilon_{abcd} \pm 2 \bar{R}_{ab})\right)
\]
Now, due to the boundary condition at the horizon Eq.(\ref{Temperature}), it is possible to single out the period \cite{Wald:1993nt,Aros:2002ub}. This yields the close expression for the entropy
\begin{equation}\label{Entropy}
  S = \hat{S} + \frac{\beta l^2}{32\pi G} \int_{\partial \Sigma_{\mathcal{H}}} I_{\xi} \omega^{ab}(\bar{R}^{cd} \varepsilon_{abcd} \pm 2 \bar{R}_{ab})
\end{equation}
where $\beta^{-1}$ is the inverse of the period and $S_0$ is an extensive constant independent of the parameter of solution.  As $\hat{S}$ cannot depend on the parameters of the solution, such as the mass, but must depends on $l$, it can be argued that $\hat{S} \sim PV$. In this case the mass must be identified with enthalpy of the system instead of the energy. For a discussion see \cite{Kastor:2009wy}.

\section{Geometry of Taub-NUT-AdS and Taub-Bolt-AdS solutions}

The four dimensional Taub-Bolt-AdS and Taub-NUT-AdS solutions are known to be described by
\begin{widetext}
\begin{equation}\label{GeneralMetric}
 ds^2 =f(r)^2(d\tau+2n\cos(\theta)d\varphi)^2 + \frac{dr^2}{f(r)^2} + (r^2-n^2)(d\theta^2+\sin(\theta)^2d\varphi^2)
\end{equation}
\end{widetext}
where
\[
f(r)^2 = \frac{(r^2+n^2)-2mr+ l^{-2}(r^4-6n^2r^2-3n^4)}{r^2-n^2}
\]
It is direct to notice that $\partial_\varphi$ and $\partial_{\tau}$ are two Killing vectors of the geometry.

The different between Taub-NUT-AdS and Taub-Bolt-AdS solutions relies on the structure of the function $f(r)^2$, the norm of the Killing vector $\partial_{\tau}$. Indeed, depending on the form of $f(r)^2$ its largest zero, say $r=r_+$, either corresponds to a point or a two dimensional surface in the geometry. This later case is called the bolt of the solution and is case of a usual stationary black hole.

The mass of this solution \cite{Araneda:2016iiy} is given by
\begin{eqnarray}
  M &=& Q\left(\partial_\tau\right) = \frac{\beta l^2}{32\pi G} \int_{\partial \Sigma_{\infty}} I_{\partial_\tau} \omega^{ab}(\bar{R}^{cd} \varepsilon_{abcd} \pm 2 \bar{R}_{ab}) \nonumber \\
  &=& m - |n|\left(1-\frac{4n^2}{l^2}\right) \nonumber\\
   &=&  \frac{(r_{+}-|n|)^2 }{2r_{+} l^2}\left((r_{+}+|n|)^2-\left(l^2-4n^2\right)\right)\label{MassOfGeneral}
\end{eqnarray}

\subsection*{Asymptotical geometry and squashed $S^3$}

It is direct to confirm that this solution is asymptotically locally AdS, and for $m=0$ corresponds to a locally AdS solution as expected. This motivated the identification of $m$ with a mass parameter in \cite{Hawking:1998ct}. The presence of the NUT charge can be noticed on the geometry of the asymptotical transverse section to the radial direction which tends to a squashed $S^3$, see for instance \cite{Johnson:2014xza}.

After redefining $\tau = 2n \psi$ in Eq.(\ref{GeneralMetric}) a suitable vierbein can be defined in terms of the dreibein $\tilde{e}^{i}$ for $S^3$ depicted in \ref{VielS3}. This veirbein is given by
\begin{equation}\label{FourDimensional}
  e^{a} = \left(\frac{dr}{f(r)}, g^1(r) \tilde{e}^1,g^2(r) \tilde{e}^2,g^3(r) \tilde{e}^3\right)
\end{equation}
with $a=0,1,2,3$. Here
\begin{eqnarray*}
  g^3 &=& 4n f(r) \\
  g^{1} = g^{2} &=& 2 \sqrt{r^2-n^2}
\end{eqnarray*}

For simplicity one can use the convention,
\[
g^i(r) \tilde{e}^i = (g^1(r) \tilde{e}^1,g^2(r) \tilde{e}^2,g^3(r) \tilde{e}^3)
\]
where the repetition of $i$ indexes does not imply summation. For the rest of work the same convention will be used. In terms of this vielbein is direct to observe, from its asymptotical form
\[
\lim_{r\rightarrow \infty }ds^2 \approx \frac{l^2 dr^2}{r^2} + r^2 \left(\left(\frac{2n}{l}\right)^2 (e^3)^2 + (\tilde{e}^1)^2 + (\tilde{e}^2)^2 \right),
\]
the presence of the NUT. Indeed, for $n = \pm l/2$ this condition disappears and the transverse section becomes a sphere.

The torsion free connection of this vielbein is given by
\begin{eqnarray}
  \omega^{0i} &=& -f(r)\frac{d g^i}{dr}  \tilde{e}^i  \nonumber\\
  \omega^{ij} &=& C^{ij}_{\hspace{2ex} k} \tilde{e}^{k}\label{TorsionFreeConnection}
\end{eqnarray}
where
\[
 C^{ij}_{\hspace{2ex} k} = \frac{(g^i)^2 + (g^j)^2 - (g^k)^2}{g^i g^j}\varepsilon_{ijk}
\]
Finally, $\bar{R}^ab$ is given by
\begin{widetext}
\begin{eqnarray}
  \bar{R}^{0i} &=& \left(-\frac{d}{dr}\left(f \frac{d g^i}{dr} \right)\frac{f}{g^i} + \frac{1}{l^2}\right) e^{0} e^{i} + f \left(C^i_{\hspace{1ex} kl} \frac{d g^k}{dr} - \varepsilon^{i}_{\hspace{1ex} kl} \frac{d g^i}{dr} \right) \tilde{e}^k \tilde{e}^l\\
  \bar{R}^{ij}  &=&  \left(\frac{g^i g^j}{l^2}-f^2 \frac{d g^i}{dr}\frac{d g^j}{dr} \right) \tilde{e}^i \tilde{e}^j + \left(C^{ij}_{\hspace{2ex} k}\varepsilon^k_{\hspace{1ex} lm} + C^{i}_{\hspace{1ex} kl} C^{kj}_{\hspace{2ex} m} \right)\tilde{e}^l \tilde{e}^m +  \frac{d }{dr} C^{ij}_{\hspace{2ex} k} dr \tilde{e}^k \nonumber
\end{eqnarray}
\end{widetext}

\section{Entropy}

As mentioned above the asymptotical conserved charges of the Taub-Bolt-AdS solution were discussed in \cite{Araneda:2016iiy}. Following \cite{Wald:1993nt}, in a matter of speaking, only rests to discuss to show that entropy can be obtained as the Noether charge associated Killing vector $\partial_\tau$ on $\partial \Sigma_{\mathcal{H}}$.

Before to proceed a general geometrical consideration can be made concerning these solutions. Following the standard approach the periods are to be fixed to avoid a conical singularity in the plane $r\psi$ at $r=r_+$. As expressed in Eq.(\ref{FourDimensional}) the period of $\psi$ is fixed such that $0\leq \psi < 4\pi$, and therefore, independently of the particular value of $r_+$, it must be satisfied that
\begin{equation}\label{TheFixingOfF}
  \left.\frac{d}{dr}(f^2(r))\right|_{r=r_{+}}=\frac{1}{2n}
\end{equation}
With this in mind, one must stress that the difference between both geometries is that while for NUT solution $r_{+}=|n|$, for bolt solution $r_{+}=r_{b}> |n|$.

For the solution above its direct to demonstrate that
\begin{eqnarray}
  I_{\partial_{\psi}} \omega^{ab} &=& -f(r)\frac{d g^3}{dr} \delta^{ab}_{01} + \frac{1}{2}\delta^{ab}_{ij} C^{ij}_{\hspace{2ex}3}\label{TheProjection}\\
  &=& -f(r)\frac{d g^3}{dr} \delta^{ab}_{01} + \left(2-\frac{(g^3)^2}{(g^2)^2} \right)\delta^{ab}_{12}\nonumber\\
   &=& -2n\frac{d f(r)^2}{dr} \delta^{ab}_{01} + 2 \delta^{ab}_{12} \left(1- 2n^2 \frac{f^2(r)}{r^2-n^2}\right)\nonumber
\end{eqnarray}

To study the behaviour near $r=r_{+}$ is necessary to separate the NUT and Bolt cases.
\subsection*{Taub-NUT}
As mentioned above the Taub Nut AdS solution has a (anti)self dual Weyl tensor. To satisfy such a condition the value of $m$ in Eq.(\ref{GeneralMetric}) must be given by
\begin{equation}\label{TaubNutMass}
 m = |n|\left(1-4\frac{n^2}{l^2}\right).
\end{equation}
It is worth to mention that the case $m=0 \leftrightarrow n = \pm l/2$ corresponds to a (locally) AdS space \cite{Chamblin:1998pz}. In this case
\begin{equation}\label{TaubNUTAdS}
f^2(r)_{\textrm{NUT}} = \left(\frac{r-|n|}{r+|n|}\right)\left(1 + \frac{(r-|n|)(r+3|n|)}{l^2}\right).
\end{equation}
where one can notice that $r_+ = n$ explicitly. This determines that the near horizon geometry is given by
\begin{eqnarray}
  \lim_{r \rightarrow n} ds^2  &\approx& 2|n| \left( \frac{dr^2}{r-|n|} + 4 (r-|n|) ds^2_{S^3}\right) \\
  &\approx& d\rho^2 + \rho^2 ds^2_{S^3}. \label{TaubNUTNearHo}
\end{eqnarray}
One can readily notice that near $r=r_{+}=n$ geometry has a $SO(4)$ symmetry and that there is no horizon as $r=\pm n$ defines a point in this geometry. Furthermore, every plane between the orbit of a SO(4) symmetry and $\rho$ becomes flat as $r\rightarrow n$.

As mentioned above, Taub-NUT-AdS has vanishing charges \cite{Araneda:2016iiy}. Unlike the usual black hole case, here the period in the limit $r\rightarrow n$ can computed from the condition in Eq.(\ref{Temperature}). This result merely determines, or confirms, that the existence of non trivial periods on $S^3$ as it shrinks to a point.

\subsection*{Taub-Bolt}

For Taub-Bolt-AdS, as mentioned above, the form of the metric is similar. However the value of $r_+ > n$. In fact,
\begin{eqnarray}
  f(r)^2_{\textrm{Bolt}} &=& \frac{1}{r^2-n^2} \left[4\,{\frac {{r}^{4}}{{n}^{2}{l}^{2}}}+{\frac {{r}^{2}}{{n}^{2}} \left(
4-24\,{\frac {{n}^{2}}{{l}^{2}}} \right) }\right. \nonumber\\
   &+& \frac {r}{n} \left[ -4\,{
\frac {{n}^{2}{s}^{3}}{{l}^{2}}}+ \left( 24\,{\frac {{n}^{2}}{{l}^{2}}
}-4 \right) s \right. \nonumber \\
& + &  \left.\left. {\frac {1}{s} \left( 12\,{\frac {{n}^{2}}{{l}^{2}}}-4
 \right) }  + 4-12\,{\frac {{n}^{2}}{{l}^{2}}}\right] \right],\label{FBolt}
\end{eqnarray}
where $s$ is parameter. See for instance \cite{Page:1985hg}. It is worth to stress that the value of $m$ in Eq.(\ref{GeneralMetric}) in this case is given by
\begin{equation}\label{TaubBoltMass}
 m =\frac{1}{2}\,{\frac { \left( {s}^{4}-6\,{s}^{2}-3 \right) {n}^{3}}{{l}^{2}s}}+
\frac{1}{2}\,{\frac { \left( {s}^{2}+ 1 \right) n}{s}}.
\end{equation}
which vanishes for $\frac{n^2}{l^2} = \frac{s^2+1}{6s^2+3-s^4}$. In this case the solution is a locally AdS space with squashing $ \frac{4(s^2+1)}{6s^2+3-s^4}$. Conversely, it must be noticed for $n=\pm l/2$, the case where the squashing disappears at the asymptotical region, $m = \frac{1}{16}{\frac { \left( s^2-1 \right) ^{2} }{s}}$. Obviously for $s=\pm 1$, in this case, the AdS space case is recovered.

Because $r_+ > n$  the near horizon geometry is described by
\[
\lim_{r\rightarrow r_+} ds^2 \approx d\rho^2 + \frac{\rho^2}{4} (d\psi + \cos(\theta)d\varphi)^2 + (r_+^2 - n^2)(d\theta^2 + \sin^2(\theta) d\varphi^2)
\]
which is the generalization of a usual stationary black hole.

To compute the temperature from Eq.(\ref{Temperature}) in this case is unnecessary as this is fixed by $f^2(r_{+})' = 1/(2n)$ for whatever value $r_+>|n|$. However, there is another aspect. After a direct computation for Taub-Bolt-AdS
\begin{equation}\label{Pro}
 \left. I_{\partial_{\psi}} \omega^{ab}\right|_{\partial \Sigma_{\mathcal{H}}} =  -\delta^{ab}_{03} + 2 \delta^{ab}_{12}.
\end{equation}
which differs in $\delta^{ab}_{12}$ from the analogous computation for Taub-NUT-AdS. This difference indicates, since the eigenvalues are identical, that the eigenvectors differ. This due to that for Taub-Bolt-AdS the orbits of the symmetries on $S^2$ do not shrink to points as $r\rightarrow r_+$, showing the presence of a genuine Killing horizon at $r_+$.

\subsection*{The entropy of Taub-Bolt}

After these previous considerations it is possible to evaluate Eq.(\ref{Entropy}) with $\beta^{-1}=8\pi n$, the temperature, yielding
\begin{equation}\label{EntropyTaubBolt}
  S_{TB} = \frac{3}{G}\, \left( {r}_{+}^{2} -{n}^{2} \right) \pi = \frac{3}{4G} A_{\mathcal{H}},
\end{equation}
with $A_{\mathcal{H}}$ the area of the horizon. The difference with the usual area 1/4 law is due to the presence of the NUT and has been noticed previously, see for instance \cite{Mann:2004mi,Astefanesei:2004ji,Johnson:2014pwa,Johnson:2014xza}. It can be noticed that for $r_{+} \rightarrow |n|$ this expression vanishes. This is expected as this corresponds to the Taub-NUT-AdS case. It is interesting to compare this computation with the discussion in \cite{Hawking:1998ct,Fatibene:2000jm}, where a careful discussion is necessary to address  the presence of the NUT charge.

\section{Free Energy}

In order to compute the value of the action principle is natural to reexpress the action principle Eq.(\ref{RPactionFinal}) in terms of fields
\[
F^{0i} = \bar{R}^{0i} \pm \frac{1}{2}\varepsilon^{0ijk}\bar{R}_{jk} \textrm{ and } F^{jk} = \bar{R}^{jk} \pm \varepsilon^{jk0i}\bar{R}_{0i}.
\]
In fact, this simplifies significatively the computations as   $F^{0i}F_{0i} = F^{ij}F_{ij}$ and thus the action principle actually can be merely rewritten as
\begin{equation}\label{NewActionPrinciple}
  I = \frac{l^2}{16\pi} \int F^{0i}F_{0i}.
\end{equation}
The next step corresponds to analyze the projection of $F^{0i}F_{0i}$ in terms of the vielbein. Through the identity \[
F^{0i}F_{0i} = 2 \tilde{e}^{3} I_{\partial_{\psi}} (F^{0i} ) F_{0i}
\]
where
\[
I_{\partial_{\psi}} F^{0i} = -dr  \frac{d}{dr}\left(-f \frac{d}{dr}g^3 \pm C^{12}_3 \right),
\]
the action principle can be written as
\begin{equation}\label{TheActionRewritten}
I =  \frac{l^2}{8\pi} \int -\tilde{e}^3 \wedge dr \frac{d}{dr}\left[ -f \frac{d}{dr} g^3 \pm C^{12}_3 \right]^{\infty}_{r_+} \wedge \left(\tilde{e}^1\wedge \tilde{e}^2 I_{\tilde{E}_1\wedge \tilde{E}_2} F_{0i} \right),
\end{equation}
where it must be noticed that
\[
\int^{\infty}_{r_+} dr \frac{d}{dr}\left[ -f \frac{d}{dr} g^3 \pm C^{12}_3 \right] = \left[ -f \frac{d}{dr} g^3 \pm C^{12}_{\hspace{2ex} 3} \right]^{\infty}_{r_+}.
\]
Moreover, it can be recognized the expression of the Noether charges, see Eq.(\ref{NotherCharges}), can be identified. Therefore,
\begin{equation}\label{TheActionOnCharges}
I =  8n\pi \left(\left.Q(\partial_\tau)\right|_{\partial \Sigma_\infty}-\left.Q(\partial_\tau)\right|_{\partial \Sigma_{\mathcal{H}}}\right)
\end{equation}
where $8n\pi$ is the inverse of the temperature. Finally, this yields
\begin{equation}\label{FreeEnergy}
  I = \frac{1}{T} M - (S-S_0) = \frac{F}{T},
\end{equation}
where $F$ can casted as the free energy of the system.

\section{Conclusions and comments}

In this article the discussion about the ground sates with nut charges in the context of the improved action principle presented in \cite{Araneda:2016iiy} is continued. In this approach one considers spaces with non trivial (anti) self dual Weyl tensor as ground states. In this case these have non vanishing NUT charges. In this article was shown that indeed this spaces can be casted as ground states as the action principle, Eq.(\ref{RPactionFinal}) provides a well defined thermodynamics in spite of the presence of a NUT charge. Moreover, it is also noteworthy that the entropy can be recovered from the expression in Eq.(\ref{EntropyTaubBolt}), basically the generalization Wald's expression.

The generalization of these results to the known extensions of the Taub-Bolt-AdS spaces is direct. Either spaces whose transverse section tends to squashed $H_3$ yet they are asymptotically locally AdS. The generalization to space with electric and angular momentum is direct as well.

Unfortunately the generalization to higher dimensions is not direct. The condition of (anti) self duality of Weyl tensor cannot be extended to dimensions higher than 4. Furthermore, the Pontryagin density in higher dimensions is not directly connected with a conformal invariant on-shell. 

\acknowledgments

This work was partially funded by grants FONDECYT 1151107,1140296, UNAB DI-735-15/R. I thank DPI20140115 for some financial support. I would like to thank R. Olea for enlightening conversations and Sourya Ray for calling my attention to his work.

\appendix

\section{A useful and non trivial vielbein on $S^3$}\label{VielS3}

In order to construct a suitable vielbein for $S^3$, defined by the quadratic form $V^2 + X^2 + Y^2 + Z^2 =1 $, one can use Euler angles such that
\begin{eqnarray*}
  X + iY &=& \cos\left( \frac{\theta}{2} \right) e^{\frac{i}{2}(\psi + \varphi)}, \\
  Z + iV &=& \sin\left( \frac{\theta}{2} \right) e^{\frac{i}{2}(\psi - \varphi)}.
\end{eqnarray*}
where $0 \leq \theta < \pi$, $0 \leq \varphi < 2\pi$ and $0 \leq \psi < 4\pi$. This defines a vielbein \cite{Eguchi:1980jx}
\begin{eqnarray}
  \tilde{e}^{1} &=& \frac{1}{2}(\sin(\psi) d\theta - \sin(\theta)\cos(\psi)) d\varphi,\nonumber\\
  \tilde{e}^{2} &=& \frac{1}{2}\left(\cos(\psi) d\theta + \sin(\theta)\sin(\psi) d\varphi \right)\textrm{ and} \label{S3VielBienExplicit}\\
  \tilde{e}^{3} &=& \frac{1}{2}(d\psi + \cos(\theta)d\varphi),\nonumber
\end{eqnarray}
that satisfies
\begin{equation}\label{S3VielBien}
  d\tilde{e}^{i} = \varepsilon^{ijk} \tilde{e}^{j} \tilde{e}^{k},
\end{equation}
where $\varepsilon^{ijk}$ is the three dimensional Levi-Civita symbol. The associated torsion free connection is given by $\tilde{\omega}^{ij} = \varepsilon^{ijk} \tilde{e}^{k}$ and thus
\begin{equation}\label{S3Curvature}
 \tilde{R}^{ij} = d \tilde{\omega}^{ij} + \tilde{\omega}^{i}_{\hspace{1ex} k} \tilde{\omega}^{kj} = \tilde{e}^{i}\tilde{e}^{j},
\end{equation}
proving that $\tilde{e}^{i}$ is indeed a proper vielbein for $S^3$. Finally, it is worth to mention that this construction yields
\begin{equation}\label{TheTransversectio}
\tilde{e}^1  \wedge \tilde{e}^2 = \frac{1}{4} \sin (\theta) d\theta \wedge d\varphi
\end{equation}
and
\[
(\tilde{e}^1)^2 + (\tilde{e}^2)^2 =  \frac{1}{4}(d\theta^2 + \sin(\theta)^2 d\varphi^2).
\]
Therefore,
\[
ds^2_{S^3} = \frac{1}{4}((d\psi + \cos(\theta)d\varphi)^2 + d\theta^2 + \sin(\theta)^2 d\varphi^2)
\]
These last results shows that this construction was made to manifestly express $S^3$ as a $S^1$ fiber bundle on a $S^2$.


\providecommand{\href}[2]{#2}\begingroup\raggedright\endgroup

\end{document}